\documentstyle[preprint,array,aps,prc,rotate]{revtex}
\begin{document}
\input{psfig}
\input{epsf}
\def\Im{\mbox{\sl Im\ }}
\def\pd{\partial}
\def\oln{\overline}
\def\olft{\overleftarrow}
\def\ds{\displaystyle}
\def\bgreek#1{\mbox{\boldmath $#1$ \unboldmath}}
\def\sla#1{\slash \hspace{-2.5mm} #1}
\newcommand{\bra}{\langle}
\newcommand{\ket}{\rangle}
\newcommand{\vep}{\varepsilon}
\newcommand{\met}{{\mbox{\scriptsize met}}}
\newcommand{\lab}{{\mbox{\scriptsize lab}}}
\newcommand{\cm}{{\mbox{\scriptsize cm}}}
\newcommand{\mcal}{\mathcal}
\newcommand{\Del}{$\Delta$}
\newcommand{\g}{{\rm g}}
\long\def\Omit#1{}
\long\def\omit#1{\small #1}
\def\beq{\begin{equation}}
\def\eeq{\end{equation} }
\def\bea{\begin{eqnarray}}
\def\eea{\end{eqnarray}}
\def\eqref#1{Eq.~(\ref{eq:#1})}
\def\eqlab#1{\label{eq:#1}}
\def\figref#1{Fig.~(\ref{fig:#1})}
\def\figlab#1{\label{fig:#1}}
\def\tabref#1{Table \ref{tab:#1}}
\def\tablab#1{\label{tab:#1}}
\def\secref#1{Section~\ref{sec:#1}}
\def\seclab#1{\label{sec:#1}}
\def\VYP#1#2#3{{\bf #1}, #3 (#2)}  
\def\NP#1#2#3{Nucl.~Phys.~\VYP{#1}{#2}{#3}}
\def\NPA#1#2#3{Nucl.~Phys.~A~\VYP{#1}{#2}{#3}}
\def\NPB#1#2#3{Nucl.~Phys.~B~\VYP{#1}{#2}{#3}}
\def\PL#1#2#3{Phys.~Lett.~\VYP{#1}{#2}{#3}}
\def\PLB#1#2#3{Phys.~Lett.~B~\VYP{#1}{#2}{#3}}
\def\PLA#1#2#3{Phys.~Lett.~A~\VYP{#1}{#2}{#3}}
\def\PR#1#2#3{Phys.~Rev.~\VYP{#1}{#2}{#3}}
\def\PRC#1#2#3{Phys.~Rev.~C~\VYP{#1}{#2}{#3}}
\def\PRD#1#2#3{Phys.~Rev.~D~\VYP{#1}{#2}{#3}}
\def\PRL#1#2#3{Phys.~Rev.~Lett.~\VYP{#1}{#2}{#3}}
\def\FBS#1#2#3{Few-Body~Sys.~\VYP{#1}{#2}{#3}}
\def\AP#1#2#3{Ann.~of Phys.~\VYP{#1}{#2}{#3}}
\def\ZP#1#2#3{Z.\ Phys.\  \VYP{#1}{#2}{#3}}
\def\ZPA#1#2#3{Z.\ Phys.\ A\VYP{#1}{#2}{#3}}
\def\half{\mbox{\small{$\frac{1}{2}$}}}
\def\quarter{\mbox{\small{$\frac{1}{4}$}}}
\def\nn{\nonumber}
\newlength{\PicSize}
\newlength{\FormulaWidth}
\newlength{\DiagramWidth}
\newcommand{\vslash}[1]{#1 \hspace{-0.5 em} /}
\def\hera{\marginpar{$\Longleftarrow$ {\scriptsize ADL}}}
\def\bela{\marginpar{$\Downarrow$ {\scriptsize ADL}}}
\def\aboa{\marginpar{$\Uparrow$ {\scriptsize ADL}}}
\def\hers{\marginpar{$\Longleftarrow$ {\scriptsize SK}}}
\def\bels{\marginpar{$\Downarrow$ {\scriptsize SK}}}
\def\abos{\marginpar{$\Uparrow$ {\scriptsize SK}}}
\tighten


\title{The equivalence theorem and the Bethe-Salpeter equation}

\author{S. Kondratyuk, A.D. Lahiff, and H.W. Fearing}
\address
{TRIUMF, 4004 Wesbrook Mall, Vancouver, British Columbia, Canada V6T 2A3}

\maketitle

\begin{abstract}
We solve the Bethe-Salpeter equation for two-particle scattering 
in a field-theoretical model using
two lagrangians related by a field transformation. The kernel of the 
equation consists of the sum of all
tree-level diagrams for each lagrangian. 
The solutions differ even if all 
four external particles are put on the mass shell, 
which implies that observables calculated by solving the
Bethe-Salpeter equation depend on the representation of
the theory. 
We point out that this violation of the equivalence theorem 
has a simple explanation and 
should be expected for any Bethe-Salpeter equation
with a tree-level kernel.
Implications for dynamical models of hadronic interactions 
are discussed.
\end{abstract}

\vspace{1cm}
Lagrangian models of hadronic interactions 

\vspace{4mm}

\vspace{4mm}
\noindent
Corresponding author: S. Kondratyuk, TRIUMF, 4004 Wesbrook Mall, Vancouver, \\
British Columbia, Canada V6T 2A3 \\
phone: +1-604-2221047 ext: 6453, fax: +1-604-2221074, e-mail: sergey@triumf.ca 

\vspace{1cm}

\section{Introduction} \seclab{intro}

It has been known for a long time \cite{Nel41}
that the same scattering matrix can be obtained  
using lagrangians related by transformations of interpolating fields.
Transformations of this kind were 
studied in detail in \cite{Sal60}.
This independence of
the S-matrix on the choice of interpolating fields is called
the equivalence theorem. 
At the root of the equivalence theorem lies
the fact that the S-matrix in quantum field theory is defined in terms of free
fields whose properties are not changed by the 
allowed transformations \cite{Haa55}. 

The relevance of the equivalence theorem to hadronic physics has been 
reemphasised recently. Utilising lagrangian models for pion-nucleon scattering
\cite{Dav96}, Compton scattering on the pion \cite{Sch95}, pion-pion and
nucleon-nucleon bremsstrahlung \cite{Fea98}, nucleon-nucleon scattering
\cite{Ada98} and 3-body scattering \cite{Fur01}, 
it has been demonstrated that various field transformations relate
lagrangians which describe completely different off-shell 3-, 4- and 
higher-point
vertices while leading to identical on-shell scattering amplitudes. Typically,
these studies involved only tree-level or one-loop
calculations and dealt with local lagrangians and field transformations.
To our knowledge there have been no multi-loop calculations illustrating the
equivalence theorem. Yet, many recent dynamical models describing pion-nucleon
\cite{Pea91} and photon-nucleon \cite{Noz90} interactions
are based on summations of infinite series of loop diagrams.
This is usually done by solving the Bethe-Salpeter equation (BSE) \cite{Sal51} 
or one of its modifications (such as 3-dimensional reductions thereof) 
with the kernel consisting of a sum of tree-level diagrams. 

It is natural to extend the work of \cite{Dav96,Sch95,Fea98} to such
multi-loop approaches and to examine the validity of 
the equivalence theorem for solutions of the BSE. 
This is the main objective of the
present letter. Due to the complexity of the equation 
a general study is very difficult, therefore we analyse the problem using 
an example of scattering of two neutral and spinless particles 
(represented by scalar 
fields $\phi$ and $\sigma$). The relative simplicity of this model helps clarify
the essential issues of the problem while avoiding many of the technical 
complications of more realistic approaches. We consider two 
representations of the model lagrangian which are related through a field 
transformation. First, we show that the equivalence theorem is satisfied
at tree level even though the lagrangian and the field transformation  
involve form factors, thereby extending conclusions of
Refs.~\cite{Dav96,Sch95,Fea98} to such 
non-local \cite{Pai50} lagrangians and transformations. 
Next, we solve the BSE for $\phi \sigma$ scattering in the two
representations. Following the usual practice, we construct
the kernel of the equation as the sum of the 
tree-level diagrams. We find that the on-shell scattering amplitudes 
calculated in the two representations are not equal to each other, 
indicating that the equivalence theorem is not obeyed by the 
solution of the BSE.

We argue that the principal origin of this representation-dependence is 
the well-known
fact that certain classes of loop graphs are not generated by the BSE with
a tree-level kernel. These loop graphs should, however, be included in the
full amplitude for which the equivalence theorem is presumed to hold.

\section{Two representations of a model field theory} \seclab{lagrs}
 
We consider a system consisting of two species of spinless neutral particles,
described by scalar fields $\phi$ and $\sigma$ whose masses are $M$ and $m$,
respectively, and assume that a $\sigma$ couples to two $\phi$'s with a strength $g$.
We take the masses 
close to the nucleon and pion masses, $M=1000$ MeV, $m=150$ MeV,
and the coupling constant close to the pion-nucleon coupling constant, $g=13$.
The interaction is equipped with a form factor, as is done usually in dynamical models
of hadronic interactions \cite{Pea91,Noz90}.
The model lagrangian reads\footnote{Throughout the calculations, 
we use the metric and conventions of \cite{Bjo64}.} 
\beq
{\mathcal{L}}={1 \over 2} (\pd \phi)^2 - \frac{M^2}{2} \phi^2 +
{1 \over 2} (\pd \sigma)^2 - {m^2 \over 2} \sigma^2 +
{g \over 2} \left[H(-\pd^2) \sigma \right] \phi
\left[ G(-\pd^2) \phi \right]  ,
\eqlab{lagr_rep1}
\eeq	    
where $\pd^2 \equiv \pd_\mu \pd^\mu$. 
A function of $(-\pd^2)$ can be viewed as a formal series  
in powers of $(-\pd^2)$. It corresponds to a form factor in momentum space.
The 3-point $\phi \phi \sigma$ vertex extracted from lagrangian \eqref{lagr_rep1} 
has the form
\beq
\frac{i g}{2} H(q^2) \left[ G(p_1^2)+G(p_2^2) \right] ,
\eqlab{vert_rep1}
\eeq
where $q$, $p_1$ and $p_2$ are the 4-momenta of the $\sigma$, 
the first and the second $\phi$'s, respectively. 

The vertex function \eqref{vert_rep1} defines the structure of the theory in the
first representation, which we will call representation (I). 
Representation (II) is introduced through a 
transformation of the $\phi$ field.
Similar to lagrangian \eqref{lagr_rep1}, we include a form factor $F$ in the transformation,
\beq
\phi  \longrightarrow  \phi + f \, \phi \left[ F(-\pd^2) \sigma \right], 
\;\;\;
\sigma  \longrightarrow  \sigma, 
\eqlab{transf}
\eeq
where $f$ is a parameter. Thus, \eqref{transf} can be regarded as a non-local \cite{Pai50}
version of transformations considered in \cite{Sal60}. 
The form factors are normalised so that $G(M^2)=H(m^2)=F(m^2)=1$. 
In the actual calculation we adopt a traditional functional dependence 
\beq
G(p^2)= \frac{M^2-\Lambda_\phi^2}{p^2-\Lambda_\phi^2} ,\;\;
H(q^2)= \frac{m^2-\Lambda_\sigma^2}{q^2-\Lambda_\sigma^2} , \;\;   
F(p^2)= \left( \frac{m^2-\Lambda_T^2}{q^2-\Lambda_T^2} \right)^2 ,
\eqlab{ff_ghf}
\eeq
where $\Lambda_\phi$, $\Lambda_\sigma$, $\Lambda_T$ are cut-off parameters and
the stronger falloff of $F$ ensures the convergence of the loop integrals 
in representation (II). 

Under the field transformation \eqref{transf} 
lagrangian \eqref{lagr_rep1} changes as
\begin{eqnarray}
 {\mathcal{L}} & \longrightarrow & \ds{ {1 \over 2} (\pd \phi)^2 - 
{M^2 \over 2} \phi^2 +{1 \over 2} (\pd \sigma)^2 - {m^2 \over 2} \sigma^2 }  
\ds{ +{g \over 2} [H \sigma] \phi [G \phi] - f [F \sigma] \phi 
\left[ (\pd^2+M^2)\phi \right] } 
\nonumber \\
 && \ds{ +{f^2 \over 2} (\pd \phi)^2 [F \sigma]^2 + {f^2 \over 2} \phi^2 
[\pd F \sigma]^2 
 + f^2 \phi (\pd \phi) [F \sigma] [\pd F \sigma] - 
{{M^2 f^2} \over 2} \phi^2 [F \sigma]^2 } 
\eqlab{lagr_rep2} \\
 && \ds{ + {{g f} \over 2} \phi [H \sigma] [F \sigma] [G \phi] + 
{{g f} \over 2} \phi [H \sigma] [G \phi F \sigma] }
\ds{ +{{g f^2} \over 2} \phi [H \sigma] [F \sigma] [G \phi F \sigma] \,,
 \nonumber }
\end{eqnarray}
where we have omitted a full derivative on the right-hand side and used a
shorthand notation in which, e.g., 
$[G \phi F \sigma] \equiv  \left[\,G(-\pd^2) (\phi [F(-\pd^2) \sigma])\,\right]$.
Lagrangian \eqref{lagr_rep2}
contains 3- ,4- and 5-point vertices which define the structure of the theory in
representation (II). The 3-point $\phi \phi \sigma$ vertex reads
\beq
\frac{i g}{2} H(q^2) \left[ G(p_1^2)+G(p_2^2) \right] +
i f F(q^2) \left[ p_1^2 + p_2^2 -2 M^2 \right]  ,
\eqlab{vert3_rep2}
\eeq
with the same definition of the 4-momenta as in \eqref{vert_rep1}. The 4-point
$\phi \phi \sigma \sigma$ vertex is
\begin{eqnarray}
&& i f^2 \left[ s+u-2 M^2 \right] F(q^{\prime 2})F(q^2) 
\nonumber \\ 
+ && \ds{ {{i g f} \over 2} } 
\Big[ H(q^2) F(q^{\prime 2}) + H(q^{\prime 2}) F(q^2) \Big]
\Big[ G(p^2) + G(p^{\prime 2}) + G(s) + G(u) \Big],
\eqlab{vert4_rep2}
\end{eqnarray}
where the 4-momenta of the incoming (outgoing) $\sigma$ and $\phi$ are
denoted as $q$ and $p$ ($q'$ and $p'$), respectively.
The usual Mandelstam variables for $\phi \sigma$ scattering are 
$s=(p+q)^2=(p'+q')^2$, $u=(p-q')^2=(p'-q)^2$.
We do not give an explicit expression for the
5-point vertex as it will not be required in the following calculations.

\Omit{
The 5-point vertex, with
two incoming $\phi$'s (4-momenta $p_1$ and $p_2$) and three incoming
$\sigma$'s ($q_1$, $q_2$ and $q_3$), has the form 
\bea
\ds{ {{i g f^2} \over 2} }  \bigg\{  \;\;\, F(q_1^2) F(q_2^2) H(q_3^2) 
\Big[ G((p_1+q_1)^2) + G((p_2+q_1)^2) + G((p_1+q_2)^2) + G((p_2+q_2)^2) \Big]
 && \nonumber \\
 + F(q_1^2) F(q_3^2) H(q_2^2) 
\Big[ G((p_1+q_1)^2) + G((p_2+q_1)^2) + G((p_1+q_3)^2) + G((p_2+q_3)^2) \Big]
 && \nonumber \\
  + F(q_2^2) F(q_3^2) H(q_1^2) 
\Big[ G((p_1+q_2)^2) + G((p_2+q_2)^2) + G((p_1+q_3)^2) + G((p_2+q_3)^2) \Big]
\bigg\}  
 && . \eqlab{vert5_rep2}
\eea
}

\section{The Bethe-Salpeter equation for $\phi \sigma$ scattering} \seclab{bse}

The $\phi \sigma$ scattering amplitude $T(q',p';q,p)$ 
can be obtained by solving the integral equation \cite{Sal51}
\bea
T(q',p';q,p) & = & V(q',p';q,p) \nonumber \\
             & + & {i \over (2 \pi)^4} \int \! d^4 k \,
V (q',p';k,p+q-k)\, S_2(p+q-k,k) \, T(k,p+q-k;q,p) \, ,
\eqlab{bse}
\eea
where $V(q',p';q,p)$ is the kernel (potential) of the equation,
$S_2(p+q-k,k)$ is a $\phi \sigma$ propagator and
the integration is done over the 4-momentum of an intermediate $\sigma$. 
The exact  
scattering amplitude (i.e.\ the one including all possible 4-point diagrams)
obeys \eqref{bse} if $S_2$ is the product of fully dressed 
$\phi$ and $\sigma$ propagators, and $V$ includes all 2-particle
irreducible 4-point graphs\cite{Sal51}. 
In this case, \eqref{bse} is sometimes called 
``the full Bethe-Salpeter equation".
Clearly, obtaining a solution of the full Bethe-Salpeter equation is not
feasible as its kernel would contain an infinite number of loop diagrams.
In practical calculations 
the kernel of \eqref{bse} is usually chosen as a sum of
lowest order diagrams and  
\beq
S_2(p+q-k,k) = D_{\phi}^{(0)}(p+q-k) D_{\sigma}^{(0)}(k)  ,
\eqlab{prop2}
\eeq
where $D_{\phi}^{(0)}$ and $D_{\sigma}^{(0)}$ are free propagators with poles
at the physical $\phi$ and $\sigma$ masses, respectively.
The resulting equation obeys two-body $\phi \sigma$ unitarity and 
is often referred to as simply 
``the Bethe-Salpeter equation" (as opposed to the full Bethe-Salpeter
equation). We will also adhere to this terminology.  

\Omit{
In this section we solve the BSE for $\phi \sigma$
scattering in the two
representations defined by lagrangians Eqs.~(\ref{eq:lagr_rep1}) and 
(\ref{eq:lagr_rep2}).
In general, the equivalence theorem implies that the
on-shell scattering amplitude does not depend on the choice of
representation. This property might be assumed to hold for the BSE as
well, thus regarding the choice of the interpolating fields 
as irrelevant for the on-shell amplitude obtained from the BSE.  
We will show, however, that the on-shell solutions of the
BSE obtained in the two representations described above differ. 
}


The tree-level amplitude equals the potential
in which the ``physical" masses and coupling constant are used. In representation (I)
it is given by the sum of s- and u-channel diagrams,   
$T^{(I)}_{\scriptsize \mbox{tree}} = T^{(I)}_{\scriptsize \mbox{tree},s} + 
T^{(I)}_{\scriptsize \mbox{tree},u}$, 
where, using \eqref{vert_rep1},  
\beq
T^{(I)}_{\scriptsize \mbox{tree},s}(q',p';q,p)=\frac{-i g^2}{4 (s-M^2)} 
\Big[ G(p^{\prime 2}) + G(s) \Big] H(q^{\prime 2}) \Big[ G(s) + G(p^2) \Big] 
H(q^2),
\eqlab{vs_rep1}
\eeq
and $T^{(I)}_{\scriptsize \mbox{tree},u}(q',p';q,p)$ can be written 
by applying the crossing transformation 
\beq
q \longleftrightarrow -q' \;\;\;
\mbox{or alternatively} \;\; p \longleftrightarrow -p' \;\;\;
(\mbox{entailing} \;\; s \longleftrightarrow u) 
\eqlab{cross}
\eeq
to $T^{(I)}_{\scriptsize \mbox{tree},s}(q',p';q,p)$.
In addition to s- and u-channel diagrams, the tree amplitude in
representation (II) contains a contact term,
$T^{(II)}_{\scriptsize \mbox{tree}} = T^{(II)}_{\scriptsize \mbox{tree},s} + 
T^{(II)}_{\scriptsize \mbox{tree},u} + T^{(II)}_{\scriptsize \mbox{tree},c}$, 
where, using Eqs.~(\ref{eq:vert3_rep2}) and (\ref{eq:vert4_rep2}),
\bea
T^{(II)}_{\scriptsize \mbox{tree},s}(q',p';q,p) & = & \ds{ {{-i} \over {s-M^2}} } 
\Big\{ f (p^{\prime 2}+s-2 M^2) F(q^{\prime 2}) + \ds{g \over 2} 
\big[ G(p^{\prime 2}) + G(s) \big] H(q^{\prime 2}) \Big\} \nonumber \\
&  \times & 
\Big\{ f (s + p^2-2 M^2) F(q^2) + \ds{g \over 2} 
\big[ G(s) + G(p^2) \big] H(q^2) \Big\}  ,
\eqlab{vs_rep2}
\eea
$T^{(II)}_{\scriptsize \mbox{tree},u}(q',p';q,p)=
T^{(II)}_{\scriptsize \mbox{tree},s}(-q,p';-q',p)$ and
$T^{(II)}_{\scriptsize \mbox{tree},c}(q',p';q,p)$ is given by \eqref{vert4_rep2}. 
The tree amplitudes in both representations are  crossing symmetric,
i.e.\ invariant under the transformation \eqref{cross}. 
This is because they comprise {\em all} diagrams dictated by the 
corresponding lagrangians at lowest order, see \figref{tree_gnrl}.

It is straightforward to verify that these lowest-order amplitudes
in the two representations  
coincide if the external particles are put on the mass shell, i.e., 
\beq
T^{(I)}_{\scriptsize \mbox{tree}}(q',p';q,p) = 
T^{(II)}_{\scriptsize \mbox{tree}}(q',p';q,p) 
\;\;\;\;\;
\mbox{if} \;\;\;\;\;
q^2=q^{\prime 2}=m^2 , \; p^2 =p^{\prime 2}=M^2  ,
\eqlab{equiv_tree}
\eeq
thus explicitly demonstrating that the equivalence theorem is obeyed at tree level. 


\Omit{
The presence of the regularising form factors $G$, $H$ and $F$ in 
lagrangians Eqs.~(\ref{eq:lagr_rep1}) and (\ref{eq:lagr_rep2})
ensures convergence of the loop integral in \eqref{bse}. We use the
monopole parametrisation of the form factors,
\beq
G(p^2)= \frac{M^2-\Lambda_\phi^2}{p^2-\Lambda_\phi^2} ,\;\;
H(q^2)= \frac{m^2-\Lambda_\sigma^2}{q^2-\Lambda_\sigma^2}   ,
\eqlab{ff_gh}
\eeq
where $\Lambda_\phi$ and $\Lambda_\sigma$ are cut-off parameters.
With $G$ and $H$ decreasing as $\sim 1/q^2$ at infinity,
an analysis of the ultraviolet behaviour of one-loop integrals in 
representation (II) shows that the required falloff of the
``transformation" form factor $F$ should be faster than $\sim 1/q^2$; 
we choose therefore
\beq
F(p^2)= \left( \frac{m^2-\Lambda_T^2}{q^2-\Lambda_T^2} \right)^2  .
\eqlab{ff_f}
\eeq
}


The Bethe-Salpeter equation \eqref{bse} iterates the tree-level potential $V$ to all orders.
As a result, in the s-channel pole diagram the $\phi$ propagator
and the $\phi\phi\sigma$ vertices become dressed whereas the non-pole
u-channel and contact diagrams are not affected \cite{Hay69}.
For this reason, in the s-channel diagram of the potential $V$
we use a bare mass $M_0$ in the $\phi$ propagator and a bare
coupling constant $g_0$ in the 3-point $\phi \phi \sigma$ vertices, 
and the physical (renormalised) quantities $M$ and $g$ in the u-channel 
and contact diagrams \cite{Pea86}. 
Thus, the off-shell potential in each representation has the form
\beq
V(q',p';q,p) = v(q',p';q,p) \, + \, 
{ \Gamma ^{(0)}(p',p'+q',q') \, \Gamma ^{(0)}(p+q,p,q) 
\over s - M_0^2} , \eqlab{potsep}
\eeq
where $v$ is the u-channel diagram (or is the sum of the u-channel
and the contact diagrams for the case of representation (II)) and
the structure of the s-channel pole
diagram is shown explicitly. The bare $\phi\phi\sigma$ vertex
$\Gamma ^{(0)}(p_1,p_2,q)$ is given by
Eqs.~(\ref{eq:vert_rep1}) (in representation (I)) or 
(\ref{eq:vert3_rep2}) (in representation (II)), with $g_0$ used instead of $g$. 

The solution of the BSE has the form
\beq
T(q',p';q,p) =  t(q',p';q,p) \, + \, 
{\Gamma (p',p'+q',q') \, \Gamma (p+q,p,q)
\over s - M_0^2 - \Sigma (s)} , 
\eqlab{schbse}
\eeq
where $t$ is the solution of the BSE with the non-pole potential $v$. The
dressed $\phi\phi\sigma$ vertex $\Gamma$ and the $\phi$ self-energy $\Sigma$  
can be written in terms of $t$,
\bea
\Gamma (p',p,q) = \Gamma ^{(0)} (p',p,q) &+ {\ds {i \over (2 \pi)^4} \int} \!
d^4 k &
\, \Gamma ^{(0)} (p',p+q-k,k) \, S_2(p+q-k,k) \nonumber \\
& & \times  \, t(k,p+q-k;q,p) ,
\eea
\beq
\Sigma (s) = -{i \over (2 \pi)^4} \int \! d^4 k 
\Gamma ^{(0)} (p+q,p+q-k,k) S_2(p+q-k,k)
\Gamma (p+q-k,p+q,k)  ,
\eeq
with the $\phi \sigma$ propagator $S_2$ given in \eqref{prop2}. 

Renormalisation imposes two requirements
on the dressed s-channel diagram of the on-shell T matrix: 
(i) it must have a pole at $s=M^2$,
and (ii) the residue at this pole must be equal to $g^2$. In other words, 
in terms of \eqref{schbse} it is required that
\beq
 \lim_{s \to M^2}   \,
\left. {\Gamma (p',p'+q',q') \, \Gamma (p+q,p,q)
\over s - M_0^2 - \Sigma (s)} \,
\right|_
{\scriptsize
{\begin{array}{c} {\scriptsize q^2=q^{\prime 2}=m^2} \\ 
{\scriptsize p^2 =p^{\prime 2}=M^2}  \end{array}} }  \, 
= \, \frac{g^2}{s-M^2}.
\eqlab{renorm}
\eeq
These two conditions are satisfied by appropriately choosing the bare mass $M_0$ and
the bare coupling constant $g_0$. The particular 
procedure used to fix $M_0$ and $g_0$ is
immaterial, as long as renormalisation conditions \eqref{renorm} 
are fulfilled (for example, one can
follow the standard renormalisation procedure described in \cite{Bjo64}).
The renormalisation conditions \eqref{renorm} are the same in both 
representations as they involve the representation-independent physical 
parameters $M$ and $g$. 

\section{Solution of the BSE in the two representations} \seclab{sol}

We solve the BSE for $\phi \sigma$
scattering in the two introduced representations. 
Through solving the BSE one
sums up a certain class of loop diagrams up to infinite order. A question arises
whether this class of loops is sufficient for the
solution to obey the equivalence theorem. As we shall see, it is not the case.
It is known that iterating the potential
according to the BSE does not generate certain loop diagrams which should 
be included in the full theory. Up to one-loop level, this is 
illustrated in 
Figs.~(\ref{fig:tree_gnrl}) and (\ref{fig:1loop_gnrl}).
At tree level, shown in \figref{tree_gnrl}, 
the scattering amplitude $T_{\scriptsize \mbox{tree}}$ contains all the diagrams 
dictated by the lagrangians in both representations. 
As a consequence, 
the choice of representation does not affect the on-shell tree
amplitude\footnote{Note that if this amplitude were not symmetric under the
crossing transformation \eqref{cross}, 
the equivalence theorem would be violated even at tree level.}. 
At one-loop level, however, the BSE generates only those diagrams which
are shown in column A of \figref{1loop_gnrl} for representation (I) and
in columns A and C for representation (II). 
We note that the diagrams not
generated in representation (I) would 
render the one-loop amplitude crossing symmetric. 
The set of diagrams missing from the BSE in representation (II) is larger: 
in addition to the graphs required by the crossing symmetry, 
also the loop correction to the 4-point vertex and the
diagrams formed from the 5-point vertex are not generated.

In Figs.~(\ref{fig:phases_mod}) and (\ref{fig:phases_f}) we compare 
the S-wave phase shifts for $\phi \sigma$
scattering obtained in representations (I) and (II). 
To ensure that the lagrangians
\eqref{lagr_rep1} and \eqref{lagr_rep2} are indeed related by the
field transformation \eqref{transf}, in both representations we kept the 
same coupling constant
$g$ and the same cut-off parameters $\Lambda_{\phi}$ and
$\Lambda_{\sigma}$. In the calculations shown we chose 
the same values for all the cut-offs, 
$\Lambda_T=\Lambda_{\phi}=\Lambda_{\sigma}=2$ GeV, and
the transformation parameter $f$ was varied between $0$ and $0.13$. 
The phase shifts obtained in representations (I) and (II)
are denoted as $\delta^{(I)}$ and $\delta^{(II)}$.  
The extent of the representation-dependence of the phase shift is related 
to the difference $\delta^{(II)} -  \delta^{(I)}$.
In \figref{phases_mod} we show the results obtained using two different values of $f$. 
With decreasing energy the representation-dependence of the phase shift gets
smaller as the loop contributions become less important.
The difference between the amplitudes in the two representations is appreciable
even at low energies, as can be seen 
by comparing the scattering lengths $a^{(I)}$ and
$a^{(II)}$ calculated in the two representations. 
We have checked that it is impossible to find transformation parameters
$\Lambda_T$ and $f$ (except for the trivial case $f=0$) such that
the phase shifts are the same in the two representations.
The dependence of the difference between the phase shifts in the 
two representations on the transformation
parameter $f$ is shown in \figref{phases_f} for three different
values of the scattering energy. 

From the above considerations it follows that the observables which
are dominated by the tree-level diagrams should not exhibit a noticeable
dependence on the choice of representation in which the BSE is solved.
In the present calculation this is exemplified by the P-wave $\sigma \phi$ 
phase shift (not shown), which 
is given almost exclusively by the tree-level kernel and is therefore 
not sensitive to the choice of representation.     

One frequently used approximation to the Bethe-Salpeter equation is 
the K-matrix approach (see, e.g., recent models \cite{Gou94}). 
Its essential simplifying feature is that the driving term,
the K-matrix, is needed only with on-shell external particles
and the scattering amplitude is obtained by algebraically iterating $K$.
The K matrix is traditionally chosen to be equal to a crossing symmetric
on-shell tree-level amplitude.   
Therefore, one should expect that the equivalence theorem holds 
for the amplitude obtained in the K-matrix approach. 
We verified that this is indeed the case. 

In the calculations presented so far, the regularisation 
form factors $G$, $H$ and $F$
were incorporated in the lagrangian and in the transformation. 
In order to check that the reached conclusions are not an artefact of this
regularisation procedure, we have solved a BSE for
$\phi \sigma$ scattering using 
the local analogues of the lagrangian \eqref{lagr_rep1} and transformation
\eqref{transf}. The kernel of the equation is obtained in both 
representations by substituting unity for 
the form factors $G$, $H$ and $F$. 
In this calculation, we chose a simple and pragmatic
way to regularise the BSE: the measure in \eqref{bse} was covariantly 
deformed as
$d^4 k \longrightarrow d^4 k \, \Lambda_M^4/(k^2-\Lambda_M^2)^2 $,
where $\Lambda_M$ is a cut-off mass.
The phase shift calculated using this model exhibits a
representation-dependence which is qualitatively similar to the one 
discussed above for the explicitly non-local lagrangians and transformations.

\section{Conclusions} \seclab{concl} 

Even though we presented our argument using a particular model,
the preceding discussion suggests that the main conclusion
of this study -- that the on-shell scattering amplitude calculated from the
Bethe-Salpeter equation is representation-dependent -- has a rather wide
validity. Indeed, the BSE does not generate the full set of loop
diagrams adequate for the fulfillment of the equivalence theorem. 
In particular, this certainly is the case for all realistic dynamical
models of hadronic interactions based on the BSE. 
The choice of a convenient representation for constructing a 
tree-level kernel influences the solution of the BSE
even on-shell, and thus should be recognised as an additional model assumption.
Our results should, however, not be construed as a claim that the equivalence 
theorem would not hold {\em in general}, i.e.\ if one were able to calculate 
the amplitude including all the loop diagrams required by the lagrangian up to
infinite order. Formally, such a calculation could be furnished as a solution of
the {\em full} Bethe-Salpeter equation which contains an infinite number of
diagrams in its kernel. The situation we have studied, using the
BSE with a tree-level kernel, is more closely related to
the type of calculations which are actually carried out.

\acknowledgments

This work was supported in part by a grant from the Natural Sciences and Engineering
Research Council of Canada.
We would like to thank Olaf Scholten for some useful comments.

           
\begin{figure}[!htb]
\centerline{{\epsfxsize 9.8 cm \epsffile[0 360 595 460]{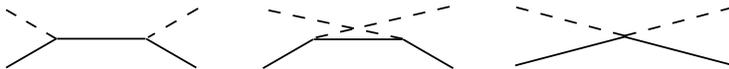}}}
\caption[f1]{
The set of diagrams included in the tree amplitude
$T_{\scriptsize \mbox{tree}}$.  
The solid and dashed lines denote $\phi$'s and 
$\sigma$'s, respectively. 
The contact term is present only 
in representation (II). Note that the 3-point vertices
are different in the two representations, see Eqs.~(\ref{eq:vert_rep1}) and
(\ref{eq:vert3_rep2}).
\figlab{tree_gnrl}}
\end{figure}

\begin{figure}[!htb]
\centerline{{\epsfxsize 13.2 cm \epsffile[15 75 575 750]{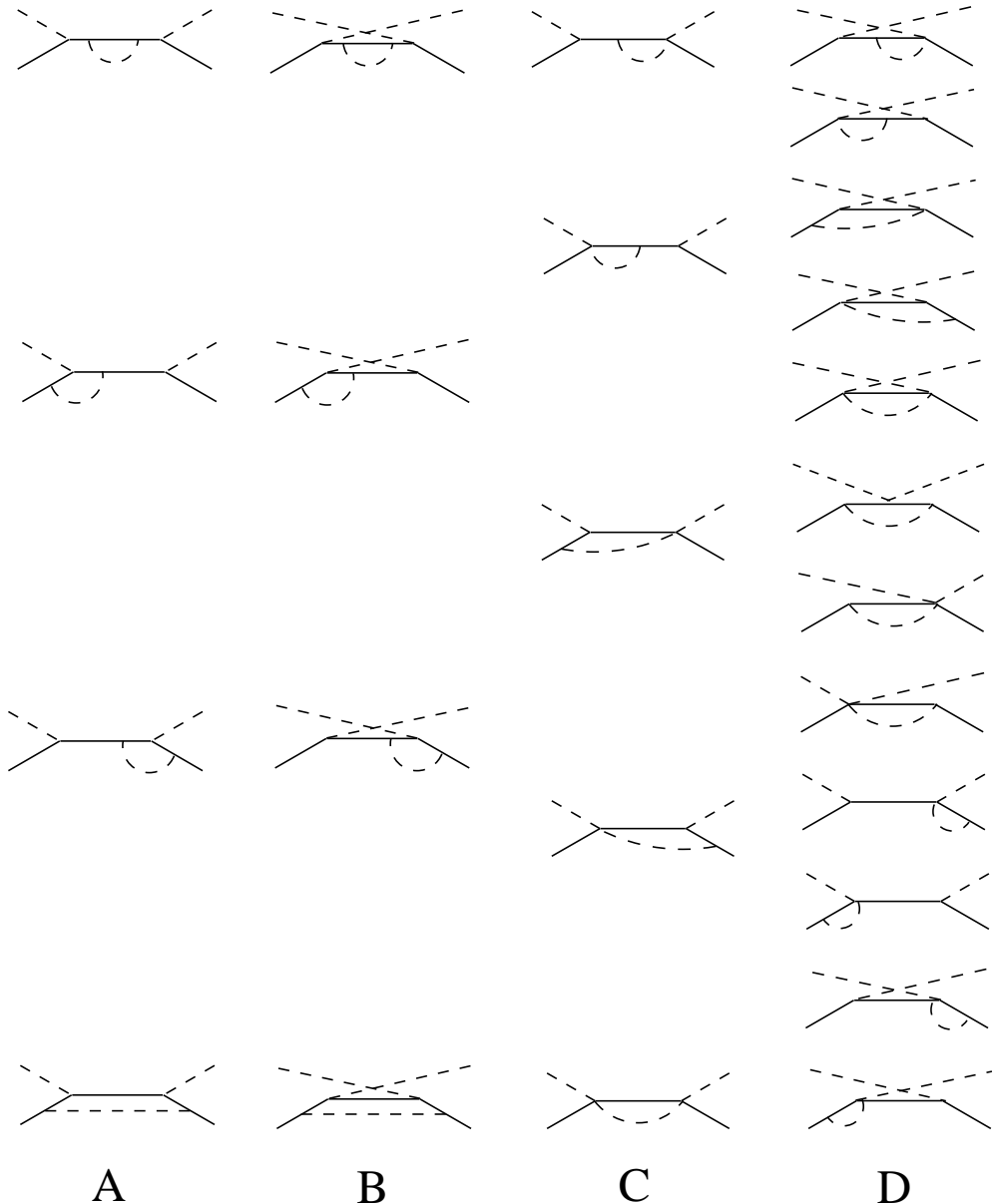}}}
\caption[f3]{
Columns A and B: 
the one-loop graphs which follow from the lagrangian in representation
(I). Columns A, B, C and D: the one-loop graphs which follow from the 
lagrangian in representation (II). 
The BSE in representation (I) (representation (II)) generates 
only the diagrams in column A (columns A and C).
Note that the 3-point vertices
are different in the two representations, see Eqs.~(\ref{eq:vert_rep1}) and
(\ref{eq:vert3_rep2}).
\figlab{1loop_gnrl}}
\end{figure}

\begin{figure}[!htb]
\centerline{{\epsfxsize 9.2cm \epsffile[35 135 555 770]{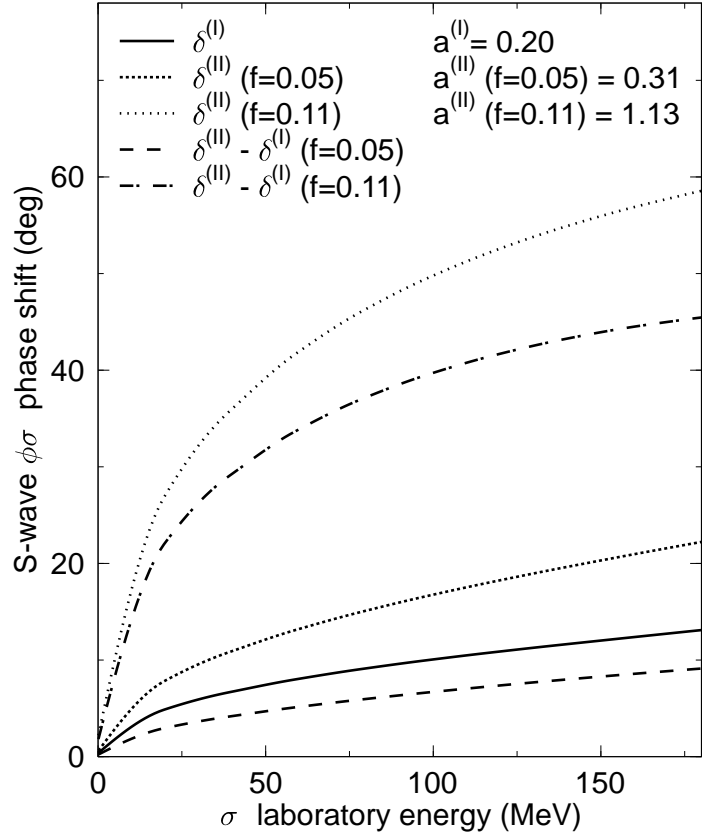}}}
\caption[f5]{
Comparison of the S-wave phase shifts for 
$\phi \sigma$ scattering obtained from the BSE in representations (I) and (II)
for two different values of the transformation parameter $f$.
The scattering lengths are given in the units of inverse $\sigma$ mass. 
The notation is explained in the text.
\figlab{phases_mod}}
\end{figure}

\begin{figure}[!htb]
\centerline{{\epsfxsize 9.2cm \epsffile[30 380 520 795]{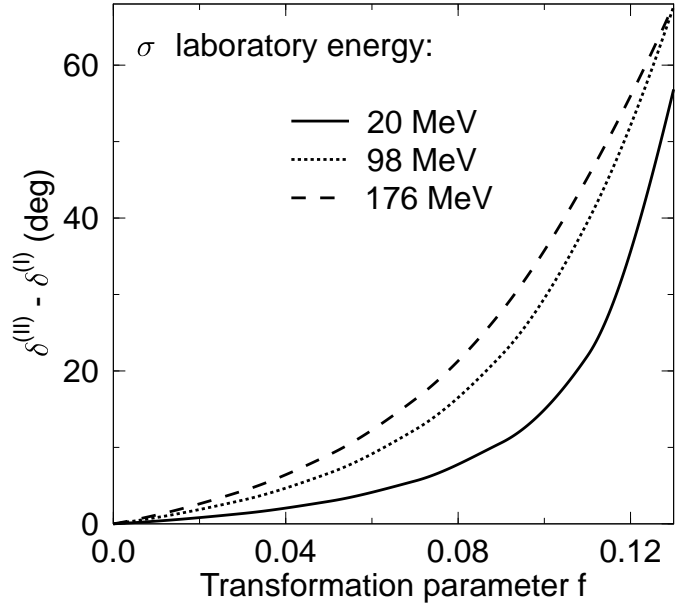}}}
\caption[f6]{
The difference between the $\phi \sigma$ S-wave phase shifts
obtained in the two representations as a function of the transformation
parameter $f$, shown for different values of
the $\sigma$ laboratory energy.  
\figlab{phases_f}}
\end{figure}




\begin{thebibliography}{99}

\bibitem{Nel41} E.C. Nelson, \PR{60}{1941}{830}; F.J.Dyson, \PR{73}{1949}{929};
                K.M. Case, \PR{76}{1949}{14}.      
\bibitem{Sal60} A. Salam, \NP{18}{1960}{681}; S. Kamefuchi \NP{18}{1960}{691};
                J.S.R. Chisholm, Nucl. Phys. {\bf 26}, 469 (1961);
                S. Kamefuchi, L. O'Raifeartaigh, and A. Salam,
                Nucl. Phys. {\bf 28}, 529 (1961);
        	S. Coleman, J. Wess, and B. Zumino, \PR{177}{1969}{2239};
		G. 't Hooft and M. Veltman, {\em Diagrammar}, 
		CERN Yellow Report 73-09, chap. 10;
		Y.-M.P. Lam, \PRD{7}{1973}{2943}; 
		M.C. Bergere and Y.-M.P. Lam, \PRD{13}{1976}{3247}.	
\bibitem{Haa55} R. Haag, Dan. Mat. Fys. Medd. {\bf 29}, no. 12, 3 (1955);
                H. Lehmann, K. Symanzik, and W. Zimmermann, 
		Nuovo Cimento {\bf 1}, 205 (1955); 
                R. Haag, Phys. Rev. {\bf 112}, 669 (1958);
                O. Greenberg, Phys. Rev. {\bf 115}, 706 (1959);
                H. Ekstein, Phys. Rev. {\bf 117}, 1590 (1960).  
\bibitem{Dav96} R.M. Davidson and G.I. Poulis, \PRD{54}{1996}{2228}.                 
\bibitem{Sch95} S. Scherer and H.W. Fearing, \PRC{51}{1995}{359}.		 
\bibitem{Fea98} H.W. Fearing, \PRL{81}{1998}{758};
                Few Body Syst.\ Suppl. 12 (2000) 263;
		H.W. Fearing and S. Scherer, \PRC{62}{2000}{034003}.		 
\bibitem{Ada98} J. Adam, F. Gross, and J.W. Van Orden, 
                Few Body Syst.\ 25 (1998) 73.
\bibitem{Fur01} R.J. Furnstahl, H.-W. Hammer, and N. Tirfessa, 
                \NPA{689}{2001}{846}.
\bibitem{Pea91} B.C. Pearce and B.K. Jennings, \NPA{528}{1991}{655}; 
                F. Gross and Y. Surya, Phys. Rev. C {\bf 47}, 703 (1993);
                C. Sch\"utz, J.W. Durso, K. Holinde, and J. Speth,
                Phys. Rev. C {\bf 49}, 2671 (1994); 
                V. Pascalutsa and J.A. Tjon, \PLB{435}{1998}{245}; 
                \PRC{61}{2000}{054003};
                A.D. Lahiff and I.R. Afnan, \PRC{60}{1999}{024608};		
		C.-T. Hung, S. N. Yang, and T.-S.H. Lee, nucl-th/0101007;
		M.F.M. Lutz and E.E. Kolomeitsev, nucl-th/0105042.
\bibitem{Noz90} S. Nozawa, B. Blankleider, and T.-S.H. Lee,
                \NPA{513}{1990}{459}; 
		Y. Surya and F. Gross, \PRC{53}{1996}{2422};
		S. Kondratyuk and O. Scholten, \PRC{64}{2001}{}, in print. 	
\bibitem{Sal51} E.F. Salpeter and H.A. Bethe, \PR{84}{1951}{1232};
	        M.M. Broido, Rep. Prog. Phys. {\bf 32}, 493 (1969). 
\bibitem{Pai50} A. Pais and G.E.Uhlenbeck, \PR{79}{1950}{145}.
\bibitem{Bjo64} J.D. Bjorken and S.D. Drell,
                {\it Relativistic Quantum Mechanics} (McGraw-Hill, 1964);
		{\it Relativistic Quantum Fields} (McGraw-Hill, 1965).
\bibitem{Hay69} R.W. Haymaker, \PR{181}{1969}{2040}.
\bibitem{Pea86} B.C. Pearce and I.R. Afnan, \PRC{34}{1986}{991}.
\bibitem{Gou94} P. F. A. Goudsmit, H. J. Leisi, E. Matsinos, B. L. Birbrair,
                and A. B. Gridnev, Nucl. Phys. {\bf A575}, 673 (1994);
                A. Yu. Korchin, O. Scholten, and R.G.E. Timmermans,
                Phys. Lett. B {\bf438}, 1 (1998);
                T. Feuster and U. Mosel, \PRC{58}{1998}{457};
		S. Kondratyuk and O. Scholten, \PRC{62}{2000}{025203}.
\end{thebibliography}
\end{document}